\documentclass[aps,prbbib,twocolumn,epsf]{revtex4}
\usepackage{graphicx}

\begin{document}
\draft

\title{Magnetic dynamics with spin transfer torques near the Curie temperature}
\author{Paul M. Haney and M. D. Stiles}

\affiliation{Center for Nanoscale Science and Technology, National
Institute of Standards and Technology, Gaithersburg, Maryland
20899-6202, USA }

\begin{abstract}
We use atomistic stochastic Landau-Lifshitz-Slonczewski simulations
to study the interaction between large thermal fluctuations and spin
transfer torques in the magnetic layers of spin valves.  At
temperatures near the Curie temperature $T_{\rm C}$, spin currents
measurably change the size of the magnetization (i.e. there is a
{\it longitudinal} spin transfer effect).  The change in
magnetization of the free magnetic layer in a spin valve modifies
the temperature dependence of the applied field-applied current
phase diagram for temperatures near $T_{\rm C}$.  These atomistic
simulations can be accurately described by a Landau-Lifshitz-Bloch +
Slonczewski equation, which is a thermally averaged mean field
theory.  Both the simulation and the mean field theory show that a
longitudinal spin transfer effect can be a substantial fraction of
the magnetization close to $T_{\rm C}$.

\end{abstract}

\maketitle

\section{Introduction}

Spin transfer torque describes the interaction between the spin of
itinerant, current-carrying electrons and the spins of the
equilibrium electrons which comprise the magnetization of a
ferromagnet.  This torque results from the spin-dependent
exchange-correlation electron-electron interaction, and leads to the
mutual precession of equilibrium and non-equilibrium spins around
the total spin.  In spin valves with sufficiently high current
density, spin transfer torque can excite a free ferromagnetic layer
to irreversibly switch between two stable configurations (typically
along an easy-axis, parallel or anti-parallel to an applied magnetic
field), or to undergo microwave oscillations.  Previous
considerations of spin transfer torque mostly focus on the {\it
transverse} response of the magnetization to spin currents
\cite{slonczewski,berger,stiles,brataas}. This is appropriate since
the temperatures used in spin valve experiments are substantially
below the Curie temperature $T_{\rm C}$ of the ferromagnets, so that
longitudinal fluctuations can be ignored. Near $T_{\rm C}$, one
expects an interplay between the large thermal fluctuations and the
nonequilibrium spin transfer torque. Generally speaking, theories of
critical phenomena in out-of-equilibrium systems have only recently
been developed \cite{mitra,feldman}, and there remain many open
questions on this topic.

Even far from the Curie temperature, temperature plays an important
role in quantitatively analyzing the dependence of the magnetic
orientation on the applied field and applied current.  The effect of
finite temperature on spin dynamics in the presence of spin transfer
torque has been modeled the macrospin approximation (fixed
magnetization length) by adding a Slonczewski torque to the Langevin
equation describing the stochastic spin dynamics \cite{li,xiao}, and
by solving the Fokker-Planck equation with the spin transfer torque
term added to the deterministic dynamics \cite{visscher}. The
Keldysh formalism provides a formal derivation of the stochastic
equation of motion \cite{nunez} for the non-equilibrium (i.e.,
current-carrying) system for a single spin of fixed magnitude. These
treatments successfully describe the thermal characteristics of
nanomagnets under the action of spin torques, such as dwell times
and other details of thermally activated switching.

For materials like GaMnAs, experiments are done near $T_{\rm C}$, so
that the {\it size} of the magnetization is substantially reduced
from its zero temperature value (temperature in Kelvin), and
undergoes sizeable fluctuations. In this case, the applicability of
a macrospin model is not clear. For field-driven dynamics, there is
theoretical work which accounts for longitudinal fluctuations near
$T_{\rm C}$ \cite{garanin1}.  This formal treatment culminates in
the construction of the Landau-Lifshitz-Bloch equation (LLB), which
is an extension of the familiar Landau-Lifshitz equation with an
additional longitudinal degree of freedom.  In this work, we
consider temperatures near the Curie temperature and include both
longitudinal fluctuations of the magnetization and the influence of
spin transfer torque.

There are a number of issues that complicate magnetic dynamics near
$T_{\rm C}$, including the temperature dependence of more basic
magnetic properties such as magnetic damping and magneto-crystalline
anisotropy, as well as the temperature dependence of the spin
transfer torque itself.  We use an atomistic approach for the
stochastic dynamics of a local moment ferromagnet with the inclusion
of spin transfer torque. Such a model is more appropriate for
systems like the dilute magnetic semiconductor GaMnAs.  Our use of
simple approximations for the temperature dependence of the magnetic
anisotropy, demagnetization field, and damping allow us to focus in
the interplay between thermal fluctuations and spin transfer torque.
We find that within this model, spin currents can change the {\it
size} of the magnetization. We give an expression for this
``spin-current longitudinal susceptibility", and propose an
experimental scheme to measure this effect.

We construct a Landau-Lifshitz-Bloch + Slonczewski (LLBS)
equation to describe both longitudinal fluctuations and spin
transfer torques. Following Ref. \onlinecite{garanin}, we verify the
applicability of the LLBS equation by comparing its results to the
atomistic results. We then analyze the LLBS equation to find the
applied field-applied current phase diagram for different
temperatures.  We find that critical switching currents are reduced
by the same mechanism exploited in heat assisted magnetic recording,
namely the temperature-induced reduction in the magnetic anisotropy
\cite{rottmayer}. We also find that regions of the phase diagram
which have been experimentally unattainable become relevant at high
temperatures. The dependence of critical currents on temperature in
these regions can provide quantitative details about the temperature
dependence of spin transfer torque.

\section{Method}

\begin{figure}[h!]
\begin{center}
\vskip 0.2 cm
\includegraphics[width=2.in]{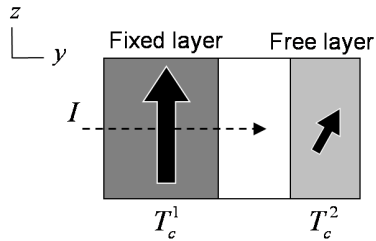}
\vskip 0.2 cm \caption{Schematic of system, two ferromagnetic layers
with different Curie temperatures.  We suppose that $T_{\rm C}^1
> T_{\rm C}^2$.}\label{fig:stack}
\end{center}
\end{figure}

To study the interplay between temperature and spin transfer torque,
we consider a spin valve with a fixed layer magnetization in the
$+\hat z$-direction with Curie temperature $T_{\rm C}^1$, and a free
layer with a smaller Curie temperature $T_{\rm C}^2$ (see Fig.
(\ref{fig:stack})). This allows for a nearly temperature independent
spin current flux incident on the free layer. We make the
approximation that all of the incoming spin current is absorbed
uniformly throughout the free layer magnetization.  This
approximation is based on two expectations. Substantial spatial and
temporal inhomogeneities in the magnetization should induce rather
irregular spatial patterns in the spin currents carried by
propagating states. This will lead to large dephasing effects, so
that the total spin current should rapidly decay away from the
interface as in the conventional picture of spin transfer
torques\cite{stiles}. In addition, in this temperature regime, and
for thin layers ($\approx 3~{\rm nm}$), magnetic non-uniformities in
the direction transverse to current flow should be more substantial
than non-uniformities {\it along} the current flow resulting from a
localized spin transfer torque.

\subsection{Stochastic Landau-Lifshitz with spin transfer}
We adopt three approaches to model the system.  The first is an
atomistic lattice model of normalized spins $\bf S$, which results
in a stochastic Landau-Lifshitz equation (SLL). We include
nearest-neighbor Heisenberg coupling with exchange constant $J$, and
an easy-axis anisotropy field of magnitude $H_{\rm an}$ in the $\hat
z$-direction.  To model the temperature dependence of the
anisotropy, we make the ansatz that the magnitude of anisotropy at
temperature $T$ is proportional to the reduced magnetization $m(T) =
M_{\rm s}(T)/M_{\rm s}^0$:
\begin{eqnarray}
 H_{\rm an}(T) = H_{\rm an}(T=0) m(T),
 \end{eqnarray}
so that the anisotropy field on spin $i$ is given by $H_{\rm
an}^i(T) = H_{\rm an}(T=0) \overline{| {\bf S}  |}S_i^z$ , where the
bar indicates a spatial average. A hard-axis anisotropy field with
magnitude $H_{\rm d}$ in the $\hat y$-direction is added to model
the demagnetization field of the thin layer.  We make an ansatz for
the form of this field to make the numerics more tractable.  We take
the demagnetization field to be uniform on all spins and given by
$H_{\rm d}^i(T) = -H_{\rm d}(T=0) \overline{ S^y } \hat y$. This
form of the hard-axis field ensures that $H_{\rm d} \sim M_{\rm
s}(T)$, and roughly captures the non-local nature of the field.
Finally, we include an applied field $H_{\rm app}$ in the $\hat
z$-direction. The Hamiltonian for spin $i$ is then:
\begin{eqnarray}
H_i &=& J \sum_{j \in {\rm n.n.}} {\bf S}_i \cdot {\bf S}_j +
\mu_{\rm B} \mu_0 \left( \frac{H_{\rm an}(T=0) \overline {|{\bf S}
|}}{2}\left(S^z_i\right)^2 \right.\nonumber \\&&~~~~~~ - H_{\rm
d}(T=0) S^y_i \left(\overline{ S^y }\right) + H_{\rm app} S_i^z
\Bigg),\label{eq:H}
\end{eqnarray}
where the sum in the first term is over nearest neighbors, $\mu_{\rm
B}$ is the Bohr magneton, and $\mu_0$ is the permeability of free
space. To model nonzero temperatures, we add damping $\alpha$ and a
stochastic field ${\bf H}_{\rm fl}$ to the equation of motion
implied by Eq. (\ref{eq:H}), with the standard statistical
properties:
\begin{eqnarray}
\langle H_{\rm fl}^\alpha(t) \rangle &=& 0, \\
 \langle H_{\rm fl}^\alpha(t) H_{\rm
fl}^\beta(t') \rangle &=& \frac{\alpha}{1 + \alpha^2} \frac{2k_B
T}{\gamma \rho } \delta_{\alpha\beta}\delta(t-t').
\end{eqnarray}
where $\alpha,\beta$ are the Cartesian components of the field,
$k_B$ is the Boltzmann constant, $\rho$ is the magnetic moment on
each lattice site, and $\gamma$ is the gyromagnetic ratio. We
numerically integrate the equation of motion using a second-order
Heun scheme \cite{palacios}. We add a Slonczewski-like spin transfer
torque term to the equation of motion for the $i$th spin, which is
given finally as:
\begin{eqnarray}
\dot{{\bf S}_i} &=& -\gamma\mu_0 \left[{\bf S}_i \times \left({\bf
H}_{\rm eff} + {\bf H}_{\rm fl}\right) -\alpha \left({\bf S}_i
\times {\bf S}_i \times {\bf H}_{\rm eff}\right) \right.\nonumber
\\&& ~~~~~~\left.+  H_{I} \left({\bf S}_i \times {\bf S}_i \times \hat
z\right)\right]. \label{eq:LLS}
\end{eqnarray}
$H_{I}$ parameterizes the spin transfer torque: $H_{I}= -I p\mu_{\rm
B} / \mu_0 e \gamma M_{\rm s}^0 \ell A$, where $I$ is the applied
current, $p$ is the spin polarization of the current, $M_{\rm s}^0$
is the zero temperature magnetization, $\ell$ is the free layer
thickness, $A$ is the transverse layer area, and $-|e|$ is the
electron charge. The effective magnetic field is given by ${\bf
H}_{\rm eff} = H_{\rm app} \hat z + H_{\rm an} \overline{| {\bf S}
|}S^z_i \hat z - H_{\rm d} \left(\overline{ S^y}\right) \hat y +
J/\left({\mu_{\rm B} \mu_0}\right) \sum_{j \in {\rm n.n.}} {\bf
S}_j$.  We use both a bulk geometry consisting of a $N=48^3$
periodic array of spins in 3 dimensions (simple cubic lattice), and
a layer geometry with an array of 100 $\times$ 100 $\times$ 15
spins.  We employ the bulk geometry in comparing the stochastic
model behavior with predictions from mean field theory, and the
layer geometry for studying the effect of spin current on
magnetization size.

\subsection{Landau-Lifshitz-Bloch + Slonczewski equation}
In the second approach, we add a Slonczewski torque term to the LLB
equation.  To derive the LLB equation, a probability distribution
for the spin orientation is assumed, which is used to find the
ensemble average of Eq. (\ref{eq:LLS}).  In addition, the nearest
neighbor exchange field is replaced by its mean-field value.  The
details of the derivation follow closely those in Ref.
\onlinecite{garanin1}, so we omit them here. The final
LLB+Slonczewski equation takes the form:
\begin{eqnarray}
\dot{{\bf m}} &=& -\gamma \mu_0 \left[\left({\bf m} \times {\bf
H}_{\rm eff} \right) + \frac{2 k_B T}{J_0 m^2} {\bf m} \cdot\left(
\alpha{\bf H}_{\rm eff}+ H_{I} \hat z\right) {\bf m} \nonumber
\right. \\&& \left.- \frac{1}{m^2}\left(1-\frac{k_B T}{J_0}\right)
{\bf m} \times {\bf m} \times \left( \alpha{\bf H}_{\rm eff} + H_{I}
\hat z\right) \right],\label{eq:LLB}
\end{eqnarray}
with an effective field given by:
\begin{eqnarray}
 {\bf H}_{\rm eff} &=& H_{\rm
app} \hat z + H_{\rm an} m^2 m_z \hat z - H_{d} m_y \hat y \nonumber
\\&& ~~~~-\frac{M_{\rm s}^0}{2\chi}\left(\frac{m^2}{m_e^2}-1\right) {\bf m}.
\end{eqnarray}
where $M_{\rm s}^0$ is the zero temperature saturation
magnetization, ${\bf m}= {\bf M}/ M_{\rm s}^0 $ is the dimensionless
magnetization with magnitude between zero and one, $m_e(T)$ is the
zero field, zero current equilibrium magnetization:
$m_e(T)=B(J_0/k_B T)$, and $B$ is the Brillouin function. $\chi(T)$
is the longitudinal susceptibility: $\chi(T)= M_{\rm s}^0
\left(\partial m_e(T)/\partial H_{\rm app}\right)$. $J_0$ is the 0th
component of the Fourier transformed exchange, and ${\bf m}$ is a
vector with size between 0 and 1.  The spin transfer torque is
parameterized by $H_I$, as described in the previous section.  The
double cross product in Eq. (\ref{eq:LLB}) is the familiar
Landau-Lifshitz damping term, which describes the relaxation of the
magnetization {\it direction} to the nearest energy minimum.  The
term longitudinal to ${\bf m}$ distinguishes the LLB equation from
the Landau-Lifshitz equation. This longitudinal term describes the
relaxation of the {\it size} of the magnetization to its steady
state value, which is determined by the temperature, applied fields,
and applied currents.

The detailed dependence of the magnetic anisotropy on temperature is
generally material specific.  In our model, the anisotropy and
demagnetization fields depend on temperature through their $m$
dependence, and vary as $m^3(T)$ and $m(T)$, respectively. The
magnetic exchange $J_0$ can also depend on temperature.  This
dependence is stronger for ferromagnets with indirect exchange
interactions (such as GaMnAs, where the magnetic interactions are
mediated by hole carriers), and weaker for local moment systems with
direct exchange (such as Fe). For simplicity we treat $J_0$ as
temperature-independent.

Finally we consider the standard Landau-Lifshitz equation with a
reduced but fixed saturation magnetization.  We find in Sec. (\ref{sec:LL})
that it is possible to appropriately modify the damping coefficient in
a standard Landau-Lifshitz approach so that the phase diagram it
predicts agrees qualitatively with those predicted by the more
complicated models.

\section{Results}

\subsection{Longitudinal spin current susceptibility}\label{sec:long}

In transition metal ferromagnets, longitudinal spin transfer, which is
another way of saying spin accumulation, is typically quite
small compared to the magnetization and has a negligible effect on the
magnetization dynamics.  However, for temperatures close to the Curie
temperatures, the longitudinal spin transfer can be a sizeable
fraction of the magnetization and can significantly affect the dynamics.

\begin{figure}[h!]
\begin{center}
\vskip 0.2 cm
\includegraphics[width=3.25in]{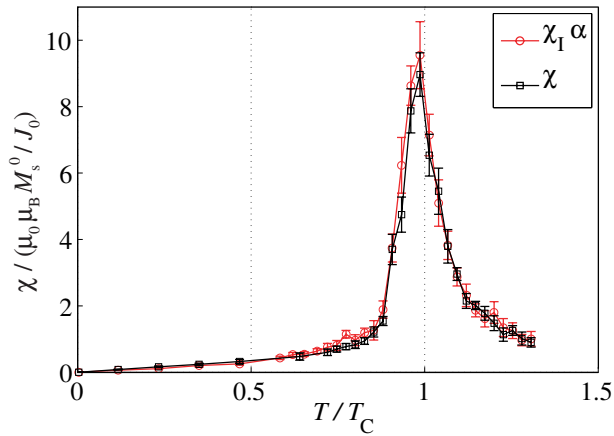}
\vskip 0.2 cm \caption{The magnetic field and spin current
susceptibility versus temperature for the stochastic Landau-Lifshitz
equation in the layer geometry.  The spin current susceptibility is
multiplied by $\alpha$.  The error bars indicate statistical
uncertainty (one standard deviation).  In the plot, $\chi$ is
rescaled by $\mu_0 \mu_B M_{\rm s}^0 /J$. }\label{fig:chi}
\end{center}
\end{figure}

Using the LLB+Slonczewski equation, it is straightforward to show
that the change in the magnetization in the presence of spin current
is
\begin{eqnarray}
\delta m(I,T) = \frac{H_{I}}{M_{\rm s}^0}\frac{\chi(T)}{\alpha}~.
\label{eq:chiI}
\end{eqnarray}
This longitudinal spin transfer effect is demonstrated in
Fig.~\ref{fig:chi}, which shows the longitudinal susceptibility to
magnetic field and spin current for a full stochastic simulation
with 100$\times$100$\times$15 spins.  (In the figure, $\chi$ is
rescaled: the magnetic field is scaled by the exchange field
$J_0/\mu_B \mu_0$, and the magnetization is scaled by $M_{\rm
s}^0$.) In the simulation, the spins' polar angle is initialized to
a uniform distribution between $\theta=0$ and $\theta=\theta_{\rm
max}$, where $\theta_{\rm max}$ is chosen so that the initial spins'
average is equal to the equilibrium value. We allow the system to
relax to steady state, and find the value of the magnetization and
its fluctuations by finding the average and standard deviation over
an interval of time (the appropriate time interval is temperature
dependent).  The fluctuations lead to the statistical uncertainty
shown in Fig. (\ref{fig:chi}).

The spin current susceptibility $\chi_I$ is defined as $\chi_I =
M_{\rm s}^0 \left(\partial m / \partial H_{I}\right)$.  We find that
$\chi$ and $\chi_I \alpha$ correspond very well, demonstrating that
Eq. (\ref{eq:chiI}) accurately describes the numerical stochastic
model.

The change in magnetization should be measurable.
The fractional change in the magnetization compared to the
zero-temperature saturation magnetization is
\begin{eqnarray}
\delta m = \left(\frac{p \mu_B }{e\gamma\mu_0\ell A \left(M_{\rm
s}^0\right)^2 } \right)\left(\frac{\chi(T)}{\alpha}\right) I
~.\label{eq:chi_dim}
\end{eqnarray}
For $T/T_{\rm C} = 0.95$, so that $\left(\chi\cdot J_0/\mu_0\mu_B
M_{\rm s}^0\right)=7$ (from Fig. (\ref{fig:chi})), and with an
exchange field of $J_0/\mu_{\rm B}\mu_0=1.2\times 10^{8} ~{\rm A/m}$
(which corresponds to a $T_{\rm C}$ of 150 K in a cubic nearest
neighbor Heisenberg model), $M_{\rm s}^0=10^6 ~{\rm A/m}$,
$I/A=10^{11} {\rm A/m^2}$, $p=0.5$, $\alpha=0.01$, and $\ell=3 ~{\rm
nm}$ gives a change compared to the zero temperature value of
$\delta m=$5 \%. Since the magnetization is reduced to approximately
20 \% of its zero temperature value at $T/T_{\rm C} = 0.95$, the
fractional change in the magnetization is approximately 25 \%.

A notable aspect of this longitudinal spin transfer is that the size
of the magnetization can either be increased or decreased according
to the direction of current flow.  For electron flow from fixed to
free layer, the free layer moment {\it increases}, while electron
flow in the opposite direction {\it decreases} the free layer
moment. This contrasts with current-induced Joule heating, which
always decreases the magnetization.

\begin{figure}[h!]
\begin{center}
\vskip 0.2 cm
\includegraphics[width=3.25in]{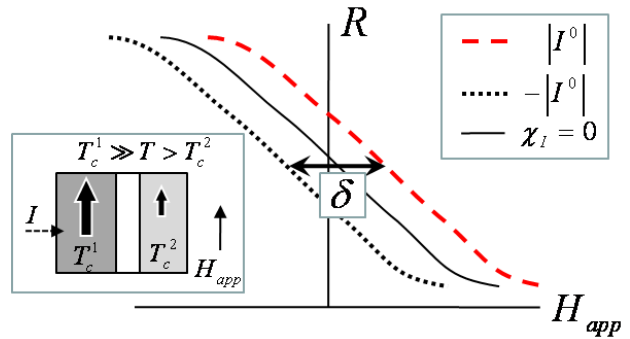}
\vskip 0.2 cm \caption{Experimental scheme for detecting longiudinal
spin transfer: for $T_{\rm C}^1 \gg T>T_{\rm C}^2$, an applied field
$H_{\rm app}$ changes the resistance $R$ via the magnetoresistance
effect. The application of a positive and negative current density
of magnitude $I^0$ shifts $m(H_{\rm app})$ in the positive and
negative direction, respectively, via longitudinal spin transfer.
The $R(H_{\rm app})$ curves therefore shift to the positive and
negative directions.}\label{fig:rh}
\end{center}
\end{figure}

This distinction can be exploited to probe the longitudinal spin
transfer by using the experimental scheme shown in Fig.
(\ref{fig:rh}). We consider the case where $T_{\rm C}^1 \gg T>T_{\rm
C}^2$.  We choose sign conventions such that a positive $H_{\rm
app}$ aligns with the fixed layer, and a positive current represents
electron flow from fixed to free layer.  In the absence of a
longitudinal spin transfer ($\chi_I$=0, black line in Fig.
(\ref{fig:rh})), the application of a magnetic field will partially
order the free layer to align or anti-align with the fixed layer.
This should cause the resistance $R$ of the device to change in some
way, according to the giant magnetoresistance effect and magnetic
order induced in the free layer (the detailed dependence of $R$ on
$H_{\rm app}$ is not important here). If a positive current $I^0$ is
applied, then the longitudinal spin transfer induces partial
ordering of the free layer, so that $m\left(H_{\rm app}=0\right) =
+\chi_I H_{I^0} /M_{\rm s}^0$. Then the curve of $m\left(H_{\rm
app}\right)$, and therefore the curve $R\left(H_{\rm app}\right)$ is
simply shifted by $+\chi_I H_{I^0}/\chi$ (the red dashed curve in
Fig. (\ref{fig:rh})). If a negative current density $-|I^0|$ is
applied, then $m\left(H_{\rm app}=0\right) = -\chi_I H_{I^0}/M_{\rm
s}^0$ and the $m\left(H_{\rm app}\right)$ and $R\left(H_{\rm
app}\right)$ curves are shifted by $-\chi_I H_{I^0}/\chi$ (black
dotted curve in Fig. (\ref{fig:rh})). This shift represents a unique
signature of longitudinal spin transfer.

Using the same parameters as before, we estimate a total shift
$\delta=2\chi_I H_{I^0} / \chi$ between $R\left(H_{\rm app}\right)$
for positive and negative current to be on the order of $8 \times
10^{5}~\rm A/m~(\approx 1~{\rm T})$. Eq. (\ref{eq:chi_dim})
indicates that materials with small exchange field (or small $T_{\rm
C}$), and those that can support large current densities show the
effect most strongly. This suggests that weak metallic ferromagnets
such as ${\rm Gd}~(T_{\rm C} = 300 ~{\rm K})$, and Fe alloys such as
${\rm Fe S_2}$ and ${\rm FeBe_5}~(T_{\rm C} = 270 ~{\rm K})$
\cite{bozorth} may be good candidates for free layer material.

\subsection{Landau-Lifshitz-Bloch-Slonczewski vs Stochastic Landau-Lifshitz} \label{sec:LLBSvsSLL}

In this section, we compare the results obtained from the full
3-dimensional stochastic LL+S equation with those obtained from the
mean-field LLBS equation.  In our numerics, we rescale time $t$ to
as $\tau = (\gamma J/\mu_{\rm B})t$, which rescales the magnetic
fields $H_{\rm eff}$ by the exchange field $H_{\rm ex} =
J/\mu_0\mu_{\rm B}$. Dimensionless fields are denoted by lowercase:
$h_{\rm app} = H_{\rm app} \mu_0\mu_{\rm B} / J$, etc.  The
dimensionless spin torque is denoted by $j_{\rm app}$, where $j_{\rm
app} = H_{I}\mu_0\mu_{\rm B} / J$. We consider a current-induced
magnetic excitation for the bulk lattice geometry at various
temperatures. The average magnetization is initialized at a
$45^\circ$ angle with respect to the $+\hat z$-direction (the
individual spins' initial direction is distributed uniformly within
$3^\circ$ in the $\theta,~\phi$ direction about
$\theta=45^\circ,~\phi=0^\circ$). The spin transfer torque is
applied to excite the magnetization away from the $\hat
z$-direction.  The parameters used are an applied field of $h_{\rm
app}=0.0001$, a demagnetization field of $h_{\rm d} = 0.01$, a
current of $j_{\rm app}=-0.0002$, and damping of $\alpha=0.1$ (the
artificially high damping was chosen to allow the numerical
simulation of the switching to be carried out in a reasonable time).
The time step used for the numerical integration is $d\tau =
0.0002$.  We vary the temperature $T$, and present results in terms
of the scaled temperature $T' = T/T_{\rm C}$.

\begin{figure}[h!]
\begin{center}
\vskip 0.2 cm
\includegraphics[width=3.25in]{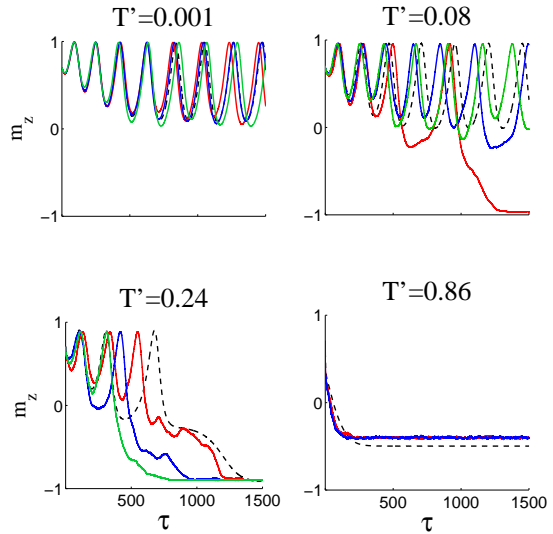}
\vskip 0.2 cm \caption{Comparison of the ($\hat z$-component)
magnetization time evolution with spin transfer torque for the
atomistic stochastic simulation and the LLB+Slonczewski equation for
various reduced temperatures $T' = T /T_{\rm C}$. The dashed line
gives the LLB+Slonczweski trajectory, while the solid lines show
various realizations of the stochastic trajectory.  The
dimensionless time $\tau$ is given by $\tau = \left(\gamma J /
\mu_{\rm B} \right) t$.}\label{fig:traj}
\end{center}
\end{figure}

As we increase temperature, we obtain trajectories of varying
complexity. Fig. (\ref{fig:traj}) compares the LLBS and several
realizations of the stochastic Landau-Lifshitz equation. For this
range of parameters, the magnetic dynamics evolves from steady
oscillations to current induced switching as the temperature is
increased. Generally, the level of correspondence between the two is
qualitatively good, although it varies between different
realizations of the stochastic dynamics. We can conclude from this
data that the LLBS equation qualitatively captures the features of
the full stochastic simulations.

The trajectories for $t=0.08$ indicate that a realization of
stochastic dynamics can exhibit the crossover from precession to
stable switching, whereas at this temperature the trajectory
obtained with the LLBS equation shows only oscillations. This
illustrates an important distinction between the stochastic
Landau-Lifshitz and LLBS models. The LLBS is an equation for the
thermally averaged magnetization, derived using an assumed
probability distribution function (in this case, a distribution
function most appropriate for temperatures well above and below
energy barriers). For this reason, the LLBS does not contain
information about fluctuations, and in particular does not capture
stochastic switching over the energy barrier.  The fluctuations may
be obtained by solving the Fokker-Planck equation, or by
supplementing the LLBS equation with stochastic fields, as done in
Ref. \onlinecite{garanin2}.

\subsection{Applied field-applied current phase diagram}

\begin{figure}[h!]
\begin{center}
\vskip 0.2 cm
\includegraphics[width=3.in]{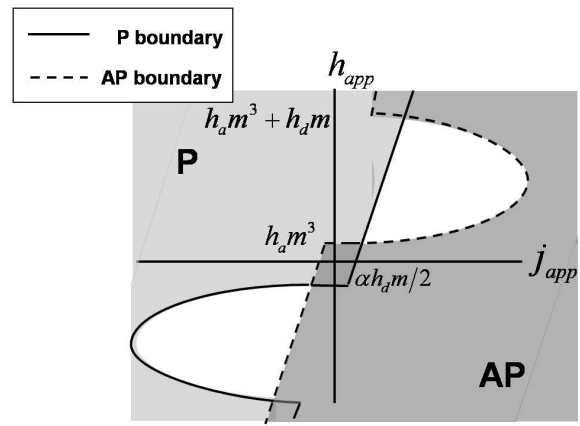}
\vskip 0.2 cm \caption{Schematic of parallel/anti-parallel stability
versus applied field and applied current.  The hysteretic box near
the origin and the fully unstable regions (white parabolic shapes)
contract in size with increasing temperature.}\label{fig:phase}
\end{center}
\end{figure}

Both high temperatures and the longitudinal degree of freedom change
the applied field-applied current phase diagram of the free magnetic
layer. Fig. (\ref{fig:phase}) shows the generic topology for regions
of stability for the parallel (``P", or $+\hat z$-direction) and
antiparallel (``AP", or $-\hat z$-direction) fixed points.  We focus
on the stability of the AP configuration for positive applied fields
(the dashed boundary in the upper-half-plane of Fig.
(\ref{fig:phase}).  We first briefly describe the main qualitative
features before providing a mathematical description. For applied
fields between $h_{\rm an}m^3$ and $h_{\rm an}m^3 + h_{\rm d}m$, the
stability boundary is a horizontal parabola, while for other values
of applied field, the stability boundary is linear with slope
$1/\alpha$.  For applied fields with magnitude less than $h_{\rm
an}m$, there is hysteresis in the current switching.  For $T=0$,
this phase diagram reduces to the known form found experimentally
\cite{ralph}.  As $T$ increases, the size of the hysteretic region
(and the switching current) decreases. Also the range of field with
the parabolic boundary decreases, and the outer edge of the parabola
gets pulled in closer to 0.  For sufficiently high temperatures,
this parabolic stability boundary should be experimentally
accessible.

A quantitative description of the phase diagram follows from Eq.
(\ref{eq:LLB}).  We determine the stability of fixed points using
the standard method of linearizing Eq. (\ref{eq:LLB}) about a fixed
point and finding parameter-dependent eigenvalues $\lambda$. A
positive real part of $\lambda$ indicates a loss of stability. This
analysis leads to the following condition for instability of the
antiparallel configuration (where it should be noted that $m$
depends on $j_{\rm app}$ through $m = m_e + \tilde{\chi}\left(h +
\frac{j_{\rm app}}{\alpha}\right)$, and $\tilde{\chi}$ is the
rescaled susceptibility, given by $\tilde{\chi}=\chi\left(
J_0/\mu_0\mu_{\rm B} M_{\rm s}^0\right)$):
\begin{widetext}
\begin{eqnarray}
{\rm Re}\left[j_{\rm app}^{\rm crit} + \alpha\left(h + h_{\rm an}
m^3 + \frac{h_{\rm d}}{2} m \frac{1-T'}{1-3T'} -
\frac{m}{2\tilde{\chi}}\left(1-\frac{m^2}{m_e^2} \right)
\frac{2T'}{1-3T'}\right) - \frac{m\sqrt{- \left( h + h_{\rm an}
m^3\right) \left( h + h_{\rm an} m^3 + h_{\rm d} m\right)}}{
1-3T'}\right]=0. \label{eq:bst1} \nonumber
\end{eqnarray}
\end{widetext}
This leads to a cubic equation for $j_{\rm app}^{\rm crit}$.
Assuming $m_e \gg \tilde{\chi}\left(h + \frac{j_{\rm
app}}{\alpha}\right)$, and expanding to 0th order in $\tilde{\chi}$
leads to an approximate, closed form for $j_{\rm app}^{\rm crit}$.
Again we distinguish between different regimes of applied field. For
$h \not\in [h_{\rm an}m^3,h_{\rm an}m^3+h_{\rm d}m]$
\begin{eqnarray}
j_{\rm app}^{\rm crit} &=& \alpha \left(h + \frac{h_{\rm d}}{2}m_e +
h_{\rm an} m_e^3 \frac{1-3T'}{1-T'} \right) \label{eq:hst1},
\end{eqnarray}
where again $m_e$ is the equilibrium magnetization in the absence of
applied field and applied current. Eq. (\ref{eq:hst1}) shows that
the slope of the boundary is temperature independent, and is given
by $1/\alpha$ (the intrinsic damping $\alpha$ is assumed to be
temperature independent).  The temperature independence of the slope
follows from the fact that the spin transfer torque increases like
$1/m(T)$, but the effective damping rate also increases as $1/m(T)$.
The intercepts of this boundary line are temperature dependent due
to the temperature dependence of $m$.  The contribution from the
easy-axis anisotropy field has an additional temperature dependence,
but the magnitude of this field is much smaller than the
demagnetization field, so it does not play an important role.  The
critical current at zero field is reduced by $m(T)$ because of the
reduction in the demagnetization field. This is important because
the demagnetization field is usually larger than applied fields, and
is therefore the primary impediment to current induced switching.
Its reduction through increased temperature offers a route to
reduced critical switching currents.

For $h_{\rm an}m^3 < h <h_{\rm an}m^3+h_{\rm d}m$, a very large spin
torque is required to stabilize the AP configuration.  The values of
current for which the AP configuration is stabilized are much higher
than those attainable experimentally, so that for this range of
fields the AP configuration is not seen \cite{BJZ}.  The approximate
critical current along the AP stability boundary is:
\begin{eqnarray}
j_{\rm app}^{\rm crit} &=&  \frac{m_e\sqrt{h(h_{\rm d} m_e-h)}}{
1-T'}.
\end{eqnarray}
The reduction in the outer boundary of the parabolic stability line
is reduced at high temperature, and this reduction can also be
traced back to the reduced magnetic anisotropy.  For low
temperatures, the application of spin transfer torques results in a
elliptical precession mostly in the easy plane about the $-\hat z$
fixed point.  To stabilize the AP configuration in this regime, the
spin transfer torque must overcome the {\it precessional} torque
(usually, the spin transfer torque must overcome the much weaker
{\it damping} torque). Assuming $h=h_{\rm d}m/2$ for definiteness,
the precessional torque decreases with $T$ as $h_{\rm d} m(T)$,
while the spin transfer torque increases like $1/m$. This implies a
value for the maximum reach of the parabola of $j_{\rm app}=
m^2(T)h_{\rm d}/(2(1-T))$. Plugging in typical values for material
parameters (the same used in Sec. (\ref{sec:long})) leads to a
critical current of $10^{12} {\rm A / m^2}$ for $T=0.95 T_{\rm C}$.
This is an order of magnitude smaller than the zero temperature
case. The behavior of this critical current versus temperature at a
fixed applied field is shown in Fig. (\ref{fig:ic}). (Solid line
gives LLBS result).  It should also be noted that the stochastic
trajectories (shown in Fig. (\ref{fig:traj})) indicate that thermal
fluctuations can effectively drive the system out of the
precessional state and into the static antiparallel configuration.

\subsection{Comparison with Landau-Lifshitz-Slonczewski}\label{sec:LL}

\begin{figure}[h!]
\begin{center}
\vskip 0.2 cm
\includegraphics[width=3.in]{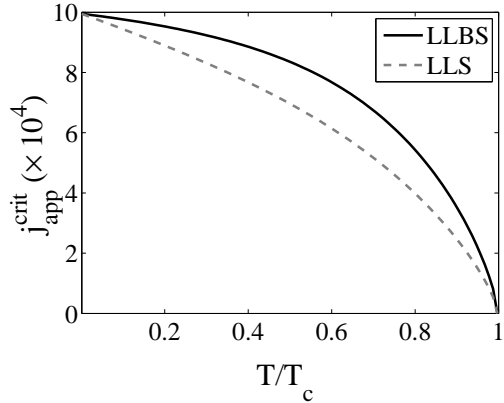}
\vskip 0.2 cm \caption{Critical current versus temperature for LLBS
and LLS equations.  The parameters are: $h_{\rm app} = -0.001$,
$h_{\rm d} = 0.01, h_{\rm an} = 0.0001$.  Recall that all fields are
scaled by the exchange field.}\label{fig:ic}
\end{center}
\end{figure}

The Landau-Lifshitz-Slonczewski (LLS) equation can be modified to
emulate the LLBS equation. Based on the qualitative behavior of the
LLBS equation, a suitable form for a temperature dependent LLS
equation for a nanomagnet of reduced magnetization size $m$ and
orientation $\hat n$ is:
\begin{eqnarray}
\dot{\hat n} = - \gamma\mu_0\left( \hat n \times {\bf H}_{\rm eff} -
\frac{\alpha}{m} \hat n \times \hat n \times {\bf H}_{\rm eff} -
\frac{H_{I}}{m} \hat n \times \hat n \times \hat z\right) \nonumber
\end{eqnarray}
where ${\bf H}_{\rm eff} = {\bf H}_{\rm app} - m H_{\rm d} n_y \hat
y + m^3 H_{\rm an} n_z \hat z$, and the temperature dependence is
contained entirely in $m(T)$.  Clearly the divergence of the damping
at $T=T_{\rm C}$ is unphysical, however a more detailed treatment of
damping near $T_{\rm C}$ is beyond the scope of this paper. The
differences between this LLS equation and the LLBS equation are
quantitative (as opposed to qualitative) in nature. One difference
is in the dependence of the critical current on temperature for
$h_{\rm an}m^3 < h <h_{\rm an}m^3+h_{\rm d}m$. Fig. (\ref{fig:ic})
shows the prediction based on the LLS equation.

The LLS equation equation neglects the longitudinal spin transfer
and applied field susceptibility, which are responsible for
dynamically changing the size of the magnetization (and therefore
the size of the effect fields) during a switching event, or other
magnetization dynamics. However, Fig. (\ref{fig:ic}) shows
qualitative agreement between the critical currents found in both
LLBS and LLS models.  This is indicative of the fact that for the
applied field-applied current phase diagram, the spin-current and
applied-field longitudinal susceptibilities play a role that is
secondary to the more pronounced effects of temperature reduced
anisotropies.

\section{Discussion}

Spin transfer torques can affect the longitudinal fluctuations of a
ferromagnet near its critical temperature.  To consider these
effects, we studied an atomistic, stochastic
Landau-Lifshitz-Slonczewski simulation at high temperatures.  We
find that there is a longitudinal spin transfer effect, and estimate
that at temperatures near $T_{\rm C}$, spin currents can measurably
change the size of the magnetization.  We then supplemented the
Landau-Lifshitz-Bloch equation with a Slonczewski torque term, and
verified that this model captures the qualitative features of the
stochastic simulations.  We showed that the applied field-applied
current phase diagram undergoes large changes in the presence of
high temperatures, and that these changes may be useful for reducing
critical switching currents and for studying the detailed behavior
of the temperature dependence of the spin transfer torque.  It
should be emphasized that these results are predicated on a
disordered local moment model of a ferromagnetic phase transition.
This model leads to an effective damping that increases with
temperature as $1/m(T)$, which effectively counteracts the similar
$1/m(T)$ increase in the magnitude of spin transfer torque.
Materials that undergo a Stoner transition should also have a
$1/m(T)$ dependence for the spin transfer torque, but a different
temperature dependence for damping.  Such materials should therefore
behave differently than the model considered here.

The experimental system relevant for the effects we describe (shown
schematically in Fig. (\ref{fig:stack})) should be relatively
straightforward to fabricate.  Jiang {\it et al.} considered a
similar system \cite{jiang}, although that work dealt with other
issues such as the ferrimagnet compensation point for magnetization
and total angular momentum.  By considering simpler ferromagnets
with different Curie temperatures, the role of temperature may be
more easily inferred.  It is of course necessary to account for
Joule heating in assessing the detailed temperature dependence of
the spin transfer torque. However recent experiments on domain wall
motion illustrates the feasibility of compensating for this effect
\cite{yamanouchi}.  On the other hand, experiments conducted at
fixed current with varying ambient temperatures and applied fields
may offer a more straightforward route to observing the longitudinal
spin transfer effect.

Many experiments done with dilute magnetic semiconductors deal with
domain wall motion, where thermal effects play an important role in
even the qualitative aspects of the domain wall
behavior\cite{yamanouchi}.  There are additional challenges
associated with extending this work from spin valves to continuous
magnetic textures.  Among these is the renormalization of the
exchange interaction associated with the coarse graining of the
magnetization, which becomes more important at higher temperatures
\cite{grinstein}. In addition, the crucial role played by the
demagnetization field in intrinsic domain wall pinning implies that
the finite temperature treatment of the demagnetization field must
also be handled more carefully.  For these reasons the spin valve
geometry may provide greater experimental control and admit a
simpler theoretical description.

\end{document}